\documentclass[12pt]{article}
\usepackage{amsmath}
\usepackage{amsfonts}
\usepackage{epsf}
\renewcommand{\bar}[1]{\overline{#1}}

\begin{document}

\begin{flushright}
\hfill{ CPT-2001/P.4201}\\ \hfill{ USM-TH-110}
\end{flushright}
\bigskip\bigskip
\centerline{\Large \bf Spin Transfers for Baryon Production in }

\centerline{\Large \bf Polarized pp Collisions at RHIC-BNL}
\vspace{22pt}
\centerline{\bf
Bo-Qiang Ma\footnote{e-mail: mabq@phy.pku.edu.cn}$^{a}$,
Ivan Schmidt\footnote{e-mail: ischmidt@fis.utfsm.cl}$^{b}$,
Jacques Soffer\footnote{e-mail: Jacques.Soffer@cpt.univ-mrs.fr}$^{c}$,
Jian-Jun Yang\footnote{e-mail: jjyang@fis.utfsm.cl}$^{b,d}$}
\vspace{8pt}
{\centerline {$^{a}$Department of Physics, Peking University,
Beijing 100871, China,}}
{\centerline {CCAST (World Laboratory),
P.O.~Box 8730, Beijing 100080, China,}}
{\centerline {and Institute of Theoretical Physics, Academia
Sinica, Beijing 100080, China}
{\centerline {$^{b}$Departamento de F\'\i sica, Universidad
T\'ecnica Federico Santa Mar\'\i a,}}
{\centerline {Casilla 110-V, 
Valpara\'\i so, Chile\footnote{Mailing address}}
{\centerline {$^{c}$Centre de Physique Th$\acute{\rm{e}}$orique,
CNRS, Luminy Case 907,}}
{\centerline { F-13288 Marseille Cedex 9, France}}
{\centerline {$^{d}$Department of Physics, Nanjing Normal
University,}}
{\centerline {Nanjing 210097, China}}
\vspace{10pt}
\begin{center} {\large \bf Abstract}
\end{center}
We consider the inclusive production of longitudinally polarized
baryons in ${\vec p}p$ collisions at RHIC-BNL, with one
longitudinally polarized proton. We study the spin transfer
between the initial proton and the produced baryon as a function
of its rapidity and we elucidate its sensitivity to the quark helicity distributions of the proton
and to the polarized fragmentation functions of the quark into the
baryon. We make predictions using an SU(6) quark spectator model
and a perturbative QCD (pQCD) based model. We discuss these
different predictions, and what can be learned from them, in view
of the forthcoming experiments at RHIC-BNL.

\vspace{2cm}
 \centerline{PACS numbers: 14.20. Jn, 13.87. Fh,
13.88.+e, 12.38.Bx, 12.39.Ki}

\newpage
\section{Introduction}

The proton spin structure has attracted a considerable interest in
the past few years, but in spite of significant theoretical and
experimental progress, a precise understanding is still far from
being satisfactory. In particular, the role played by antiquarks
and gluons in the nucleon spin remains unsettled, so there is a
bad need for more new data. Polarized deep inelastic scattering
(DIS) experiments at CERN, DESY, JLab and SLAC will certainly
continue helping us to gain some insight into this problem, but we
also expect a lot to be achieved by means of the Relativistic
Heavy Ion Collider (RHIC) at BNL. This facility, which has been
turned on recently, will operate several weeks a year as a
polarized $pp$ collider with high luminosity and with a center of
mass energy $\sqrt{s} = 500~ \rm{GeV}$, perhaps even
higher~\cite{Sai}. A vast spin physics program will be undertaken
at RHIC-BNL~\cite{Sai,BSSV}, which will focus not only on the
proton spin but also on inclusive production of hadrons in order
to study the hadronization mechanism and the quark  fragmentation
functions $D^H_q(z)$. Here $z$ stands for the momentum fraction of
the parent quark $q$ carried by the produced hadron $H$. In the
case of baryon production, the measurement of the baryon
longitudinal polarization allows to study spin-dependent
fragmentation functions $\Delta_L D^H_q(z)$. They contain
information on how the spin of the parent polarized quarks is
transferred to the baryon, and in turn this gives new insight into
the baryon spin structure, which is even more poorly known in the
case of the strange baryons. Among these hyperons, $\Lambda$
baryons are specially well suited for polarization studies due to
their self-analyzing property in the dominant weak decay channel
$\Lambda \to p \pi^-$. The unpolarized $\Lambda$ fragmentation
functions $D^\Lambda_q(z)$ are reasonably well determined from the
measurement of the rates in $e^+e^-$ annihilation in the energy
range $14 \leq \sqrt{s} \leq 91.2~\rm{GeV}$. However the poor
accuracy of the LEP data on the polarization of the $\Lambda$'s
produced at the $Z$ pole does not allow a unique determination of
the corresponding fragmentation function $\Delta_L
D^\Lambda_q(z)$~\cite{FSVa}. This situation has motivated a recent
study of the rapidity distribution of the spin transfer in the
reaction $\vec{p} p \to {\vec{\Lambda}} X$ at
RHIC-BNL~\cite{FSVb}, which seems to provide a good tool to
discriminate between various sets of polarized fragmentation
functions compatible with the LEP data. The present work lies
along the same lines, but we also generalize it to all the other
octet baryons, and we use different sets of polarized
fragmentation functions. We carefully analyze the sensitivity of
the various partonic subprocesses which lead to the final baryon,
and we try to clarify what can be learned  from these spin
transfer measurements. The paper is organized as follows: in the
next section we recall the basic kinematics and we make a detailed
analysis of the subprocesses in order to identify the properties
of the dominant ones. In section 3 we give a short review of the
theoretical framework which we use to construct the polarized
fragmentation functions of the octet baryons. Sec.~4 is devoted to
the $\Lambda$ baryon which deserves special attention, and we
compare our results to those from other theoretical
works~\cite{FSVb,BLTh}. Sec.~5 contains our predictions at
RHIC-BNL for all the other octet baryons. In section 6, based on
the dominant subprocess, we try to describe the spin transfers
with approximate formulae and elucidate their sensitivity to the
helicity distributions of the proton and to the fragmentation
functions of the produced baryon. Finally we give our discussion
and summary in Section 7.

\section{Kinematics and subprocesses analysis}

Let us consider the reaction $\vec{p}p\to \vec {B}X$, for the
single inclusive production of a polarized baryon $B$ of energy
$E$ (or rapidity $y$, with $y>0$ if $B$ is in the direction of the
polarized proton) and transverse momentum $p_T$. Here we assume
that both the initial proton and the final baryon are
longitudinally polarized. The spin transfer for this reaction is
defined as
\begin{equation}
 A^{B}={\sigma (s_p ,s_B)-\sigma(s_p ,-s_B) \over\sigma
(s_p ,s_B)+\sigma(s_p ,-s_B) }
\label{A}
\end{equation}
where $\sigma(s_p,s_B)=E_B d\sigma/d^3p_B$ stands for the
invariant cross section, and $s_p$,$s_B$ are the proton and baryon
spin vectors. $A^B$ is usually written as
$A^B={\Delta\sigma/\sigma}$, and~\cite{BLTh}
\begin{multline}
\Delta \sigma \equiv{E \Delta d^3\sigma \over dp^3}= \sum \limits_{abcd}
\int_{\bar x_a}^1 dx_a
\int_{\bar x_b}^1 dx_b \Delta f_a^p(x_a,Q^2)f_b^p(x_b,Q^2) \Delta D_c^B(z,Q^2) \\
{1 \over \pi z}{\Delta d\hat \sigma \over d\hat t}(ab \to cd)~,
\label{Dsig}
\end{multline}

with
\begin{equation}
\bar x_a={x_Te^y \over 2-x_Te^{-y}}\ ,\ \bar
x_b={x_ax_Te^{-y}\over 2x_a-x_Te^y}\ ,\  z={x_T\over 2x_b} e^{-y}
+ {x_T \over 2x_a}e^y \ ,
\label{xz}
\end{equation}
where $x_T=2p_T/\sqrt s$, $\sqrt s$ is the center of mass
energy of the $pp$ collision,  and $\hat{t}=-x_a p_T \sqrt{s} e ^ {-y} /z$
is the Mandelstam variable at the parton level.
The sum is running over all possible
leading order subprocesses $\vec{a}b \to \vec{c}d$  whose spin transfer is defined analogously to $A^B$,
with $\Delta d\hat\sigma/d\hat t$ in the numerator and
$d\hat\sigma/d\hat t$ in the denominator. These quantities are
known and their explicit expressions can be found in
Refs.~\cite{BRST,SV}. The $\Delta f^p$ ($f^p$) are the usual
(un)polarized parton distributions of the proton and
\begin{equation}
\Delta D_ c^B(z,Q^2) \equiv D_{c(+)}^{B(+)}(z,Q^2) -
D_{c(+)}^{B(-)}(z,Q^2)
\label{DD}
\end{equation}
describes the fragmentation of a longitudinally polarized parton
$c$ into a longitudinally polarized baryon $B$.
$D_{c(+)}^{B(\pm)}(z,Q^2)$ are the probabilities for finding a
baryon $B$ with positive or negative helicity in the parton $c$
with positive helicity, and $D_c^B(z,Q^2)$ is the sum of them.
 The variable $Q^2$ which occurs
in the parton distributions and in the fragmentation functions is
taken to be $Q^2=p_T^2$. Finally the denominator of $A^B$, which
is the unpolarized cross section $\sigma$, has a similar
expression to (\ref{Dsig}), with all $\Delta$'s removed. In the
numerical calculation of the spin transfer, $p_T$ will be
integrated with a minimal cutoff value of ${p_T}_{min}= 13
~\rm{GeV}$. In order to study the sensitivity to the $B$
fragmentation functions, we need to understand the dynamical
mechanism at work in this inclusive production. Among the numerous
channels which are involved in the summation in Eq.(\ref{Dsig}),
only three subprocesses contribute significantly to the cross
section. The  dominant subprocess is $qg \rightarrow qg$, which
has a gluon $g$ and a quark $q$ in the initial and final states,
and next we find $qq \rightarrow qq$ and $qq' \rightarrow qq'$,
where the quarks carry different flavors. In Fig.~\ref{Fig1} and
Fig.~\ref{Fig2},  we show
the contributions of these three channels to $\sigma$ and $\Delta
\sigma$ as a function of $y$, respectively. In order to stress the
role of the fragmentation functions, first the curves in Fig.~\ref{Fig1}(a)
and Fig.~\ref{Fig2}(a) are produced by setting $D_c^B=\Delta D_c^ B= 1$
(Here $B$ is $\Lambda$), and then the  curves in Fig.~\ref{Fig1}(b) and
Fig.~\ref{Fig2}(b) are given with the fragmentation functions in the pQCD
counting rules analysis, as explained below (see section 3.2). In
Fig.~\ref{Fig3}, the ratios of polarized to unpolarized cross sections are
shown for the three most important subprocesses.

We observe that, while the three contributions to the unpolarized
cross section are symmetric in {\it y}, as expected, this is not
the case for the corresponding contributions to the polarized
cross section. The negative {\it y} region is strongly suppressed
for the channels $qq \rightarrow qq$ and $qq' \rightarrow qq'$
because, as shown in Fig.~\ref{Fig3}, $\Delta \hat \sigma / \hat \sigma$ is
much smaller for $y<0$ than for $y>0$. In the case of the dominant
channel $qg \rightarrow qg$, $ \Delta \hat \sigma / \hat \sigma =
1 $ for all {\it y}, but the asymmetry is partly due to the fact
that we have assumed $D_g^B(z) = \Delta D_g^B(z)=0$ at the initial
energy scale as a first approximation (see section 6 for a
detailed discussion on the {\it y} dependence of $A^B$).

\begin{figure}
\begin{center}
\leavevmode {\epsfysize=12cm \epsffile{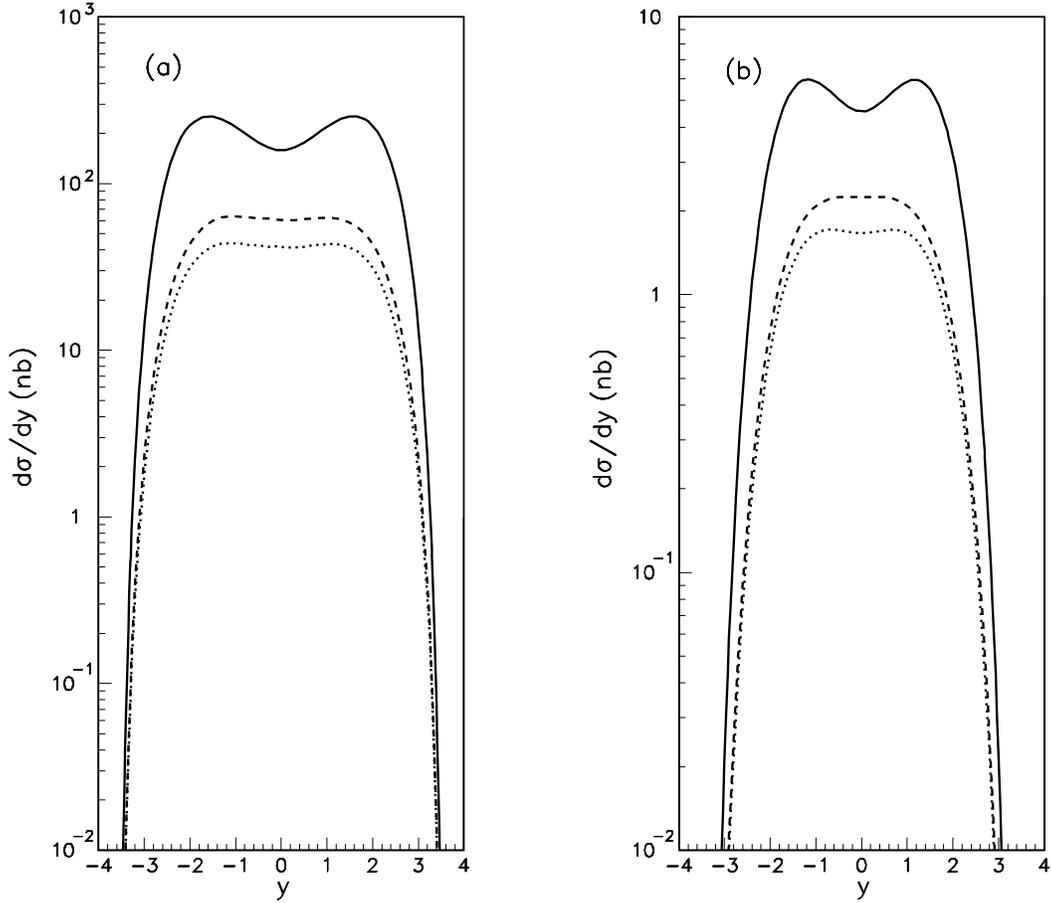}}
\end{center}
\caption[*]{\baselineskip 13pt Contributions from the three most
important channels (solid curve $qg \rightarrow qg$, dashed curve
$qq' \rightarrow qq'$ and dotted curve $qq \rightarrow qq$) to the
inclusive $\Lambda$ production unpolarized  cross sections, at
$\sqrt{s}=500~\rm{GeV}$. For comparison, the curves in (a) and (b)
are produced by setting $D_c^{\Lambda}=1$ and by using the
$D_c^{\Lambda}$ obtained in the pQCD counting rules analysis,
respectively.}
\label{Fig1}
\end{figure}

 \begin{figure}
 \begin{center}
 \leavevmode {\epsfysize=12cm \epsffile{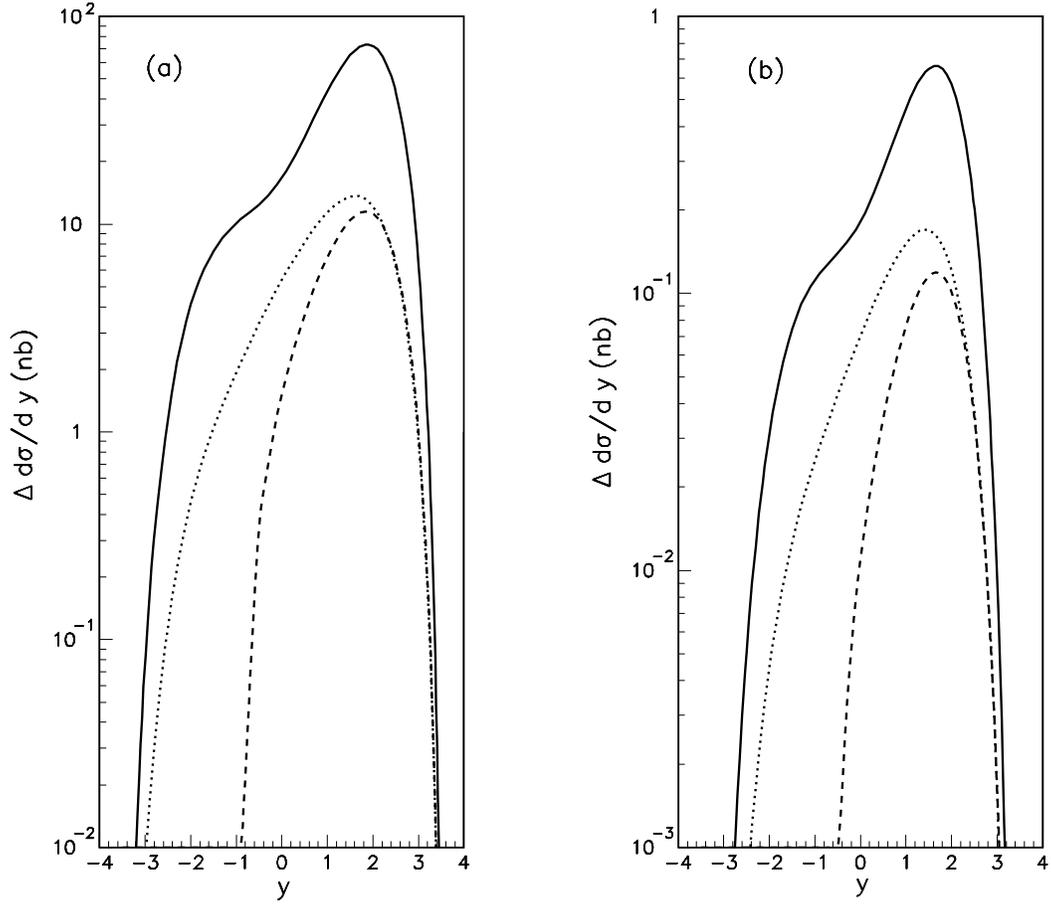}}
 \end{center}
 \caption[*]{\baselineskip 13pt
 The same labels as in Fig.~\ref{Fig1}, but for the polarized cross sections.
 }\label{Fig2}
 \end{figure}

\begin{figure}
\begin{center}
\leavevmode {\epsfysize=5cm \epsffile{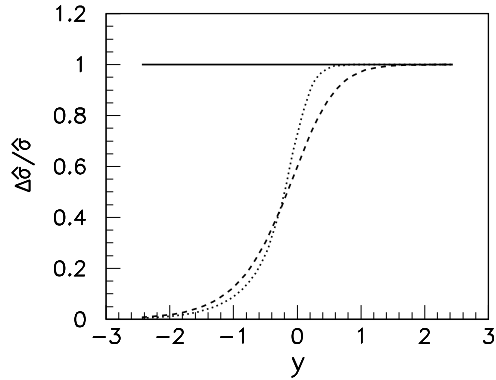}}
\end{center}
\caption[*]{\baselineskip 13pt $\Delta \hat {\sigma} /\hat
{\sigma}$ corresponding to  $x_a=x_b=0.3$ and $p_T=13 ~\rm{GeV}$
for the three most important channels, with the same labels as in
Fig.~\ref{Fig1}. }\label{Fig3}
\end{figure}

\newpage

\section{Theoretical framework for $q \to \Lambda$ fragmentation}

There have been many efforts to relate the $\Lambda$ polarization
to its spin structure. In most of these analyses, polarized
$\Lambda$ fragmentation functions were
proposed~\cite{FSVa,BLTh,Kot98} based on a simple ansatz such as
$\Delta D_{q}^{\Lambda}(z) =C_{q}(z) D_{q}^{\Lambda}(z)$ with
$C_q(z)$ either $z^{\alpha}$ or constant coefficients, or Monte
Carlo event generators without a clear physics motivation.
Therefore there is a real need to give more realistic predictions
of the $\Lambda$-polarization for future experiments such as those
at HERMES and at RHIC-BNL. In order to give more reliable
predictions of the spin transfer for the produced $\Lambda$ in
$\vec{p}p$ collisions, we employ an SU(6) quark-spectator-diquark
model and a perturbative QCD (pQCD) based counting rules analysis
since they have clear physics motivations, to describe the
polarized quark distributions in the $\Lambda$. In the following
subsections, we review  these two models for the spin structure of
the $\Lambda$ and explain how to relate the quark fragmentation
functions to the corresponding quark distribution functions.

\subsection{SU(6) quark-diquark spectator model}

The  model~\cite{Ma96,MSSY} starts from the three quark  SU(6) quark model wavefunction of the
$\Lambda$,
\begin{equation}
|\Lambda^{\uparrow} \rangle =\frac{1}{\sqrt{12}} [(u^{\uparrow} d
^{\downarrow} + d^{\downarrow} u ^{\uparrow}) -(u^{\downarrow} d
^{\uparrow} + d ^{\uparrow} u ^{\downarrow} )] s^{\uparrow} +
(\mathrm{cyclic ~~permutation}). \label{SU6}
\end{equation}
If any of the quarks is probed,  the
other two quarks can be regarded as a  diquark state  with spin 0
or 1 (scalar and vector diquarks), {\it i.e.}, the diquark serves as an
effective particle, called the spectator. In terms of quark and diquark
states, the  wavefunction of the $\Lambda$ can be rewritten  as
\begin{multline}
|\Lambda^{\uparrow} \rangle = \frac{1}{\sqrt{12}}
[V_0(ds) u^{\uparrow}-V_0(us) d^{\uparrow} -\sqrt{2} V_{+}(ds)
 u^{\downarrow} +\sqrt{2} V_{+}(us) d^{\downarrow} \\
 + S (ds) u^{\uparrow}
 +S (us) d^{\uparrow} - 2 S(ud) s^{\uparrow}]~,
 \label{WFL}
\end{multline}
where $V_{s_z}(q_1 q_2)$ stands for a $(q_1 q_2)$ vector diquark
Fock state with third spin component $s_z$, and $S(q_1 q_2)$
stands for a $(q_1q_2)$ scalar diquark Fock state. In this model,
some non-perturbative effects  between the two spectator quarks or
other non-perturbative gluon effects in the hadronic debris can be
effectively taken into account by the mass of the diquark
spectator. The model prediction of positive polarizations for the
$u$ and $d$ quarks inside the $\Lambda$ at $x \to 1$ has been
found to be  supported by the available experimental
data~\cite{MSSY}. According to the wavefunction of the $\Lambda$
in (\ref{WFL}), the unpolarized and polarized valence quark
distributions ($u_v(x), s_v(x)$ and $\Delta u_v(x), \Delta
s_v(x)$) of the $\Lambda$ can be expressed as
\begin{eqnarray}
&&u_{v}(x)=\frac{1}{4}a_V(x)+\frac{1}{12}a_S(x);\nonumber\\
&&s_{v}(x)=\frac{1}{3}a_S(x), \label{us}
\end{eqnarray}
and
\begin{eqnarray}
&&\Delta u_{v}(x)=-\frac{1}{12}\tilde{a}_V(x)+\frac{1}{12}\tilde{a}_S(x);\nonumber\\
&&\Delta s_{v}(x)=\frac{1}{3}\tilde{a}_S(x), \label{usp}
\end{eqnarray}
respectively, where $a_D(x)$ ($D=S$ for scalar spectator or $V$ for axial vector
spectator) can be expressed  in terms of the light-cone momentum
space wave function $\varphi (x, \vec{k}_\perp)$ as
\begin{equation}
a_{D}(x) \propto  \int [\rm{d}^2 \vec{k}_\perp] |\varphi (x,
\vec{k}_\perp)|^2, \hspace{1cm} (D=S \hspace{0.2cm} or
\hspace{0.2cm} V)
\end{equation}
which is normalized such that $\int_0^1 {\mathrm d} x a_D(x)=3$
and denotes the amplitude for quark $q$ to be scattered while the
spectator is in the diquark state $D$. The amplitude  for the
quark spin distributions including the Melosh-Wigner  rotation
effect~\cite{MSS} reads
\begin{equation}
\tilde{a}_{D}(x) \propto  \int [\rm{d}^2 \vec{k}_\perp]
\frac{(k^+ +m_q)^2-{\vec{k}}^2_{\perp}}
{(k^+ +m_q)^2+{\vec{k}}^2_{\perp}}
|\varphi (x,
\vec{k}_\perp)|^2, \hspace{1cm} (D=S \hspace{0.2cm} or
\hspace{0.2cm} V)
\end{equation}
with $k^+=x {\cal M}$ and ${\cal
M}^2=\frac{m^2_q+{\vec{k}}^2_{\perp}}{x}+\frac{m^2_D+{\vec{k}}^2_{\perp}}{1-x}$,
where $m_D$ is the mass of the diquark
spectator.
 In our numerical analysis, we employ the Brodsky-Huang-Lepage (BHL)
 prescription~\cite{BHL} of the light-cone momentum space wave function
 of the quark-spectator
\begin{equation}
\varphi (x, \vec{k}_\perp) = A_D \exp \{-\frac{1}{8\alpha_D^2}
[\frac{m_q^2+\vec{k}_\perp ^2}{x} +
\frac{m_D^2+\vec{k}_\perp^2}{1-x}]\},
\end{equation}
with the parameter $\alpha_D=330~ \rm{MeV}$. We take the quark masses as
$m_u=m_d=330~ \rm{MeV} $ and $m_s = 480~ \rm{MeV} $. We choose the diquark
masses  $m_{S}= 600 ~\rm{MeV} $ and $m_V= 800 ~\rm{MeV} $ for non-strange
diquark states, $m_{S}= 750~ \rm{MeV} $ and $m_V= 950~ \rm{MeV} $
for diquark states $(qs)$ with $q=u,~ d$.

\subsection{pQCD counting rules analysis}

The pQCD counting rules analysis has been successfully  used to describe
the spin structure of the nucleon~\cite{Bro95}.
The typical characteristic of the pQCD counting rules analysis lies in
that it predicts ``helicity retention", which means that the helicity of a
valence quark will match that of the parent nucleon. Explicitly,
the quark distributions of a hadron $h$ have been shown to satisfy
the counting rule~\cite{countingr} for the large $x$ region,
\begin{equation}
q_h(x) \sim (1-x)^p, \label{pl}
\end{equation}
where
\begin{equation}
p=2 n-1 +2 \Delta S_z.
\end{equation}
Here $n$ is the minimal number of the spectator quarks, and
$\Delta S_z=|S_z^q-S_z^h|=0$ or $1$ for parallel or anti-parallel
quark and hadron helicities, respectively~\cite{Bro95}.
We extend the pQCD analysis from the nucleon  case to the $\Lambda$.
More specifically, we adopt the canonical form for the quark
distributions,
\begin{equation}
\begin{array}{cllr}
&q^{\uparrow}_{i}(x)=\frac{\tilde{A}_{q_{i}}}{B_3}
x^{-\alpha}(1-x)^3+\frac{\tilde{B}_{q_{i}}}{B_4}
x^{-\alpha}(1-x)^4;\\
&q^{\downarrow}_{i}(x)=\frac{\tilde{C}_{q_{i}}}{B_5}
x^{-\alpha}(1-x)^5+\frac{\tilde{D}_{q_{i}}}{B_6}
x^{-\alpha}(1-x)^6.
\end{array}
\label{case3}
\end{equation}
with  $q_1=s$ and $q_2=u$ or $d$, where  $B_n$ is the
$\beta$-function defined by $B(1-\alpha,n+1)=\int_0^1
x^{-\alpha}(1-x)^{n} {\mathrm d} x$. Here $\alpha=1/2$
because the small $x$ behavior is controlled by the Regge
exchanges for non-diffractive valence quarks. The helicity
retention for the quark distributions in the $\Lambda$ implies
that $u(x)/s(x) \to 1/2$ and $\Delta q(x)/q(x) \to 1$ (for $q=u$,
$d$, and $s$) for $x \to 1$, and therefore the flavor structure of
the $\Lambda$ near $x=1$ is a region in which accurate tests of
pQCD can be made. There are five constraint conditions due to the
numbers of quarks,
 the quark contributions
 to the spin of the $\Lambda$, and the helicity retention property,
 \begin{equation}
 n_{u}^\uparrow + n_{u}^\downarrow=1, \label{con1}
 \end{equation}
\begin{equation}
 n_{s}^\uparrow + n_{s}^\downarrow=1, \label{con2}
 \end{equation}

\begin{equation}
 n_{u}^\uparrow - n_{u}^\downarrow = \Delta U, \label{con3}
 \end{equation}

\begin{equation}
 n_{s}^\uparrow - n_{s}^\downarrow =\Delta S, \label{con4}
 \end{equation}
and
\begin{equation}
\frac{\tilde{A}_{u}}{\tilde{A}_{s}}=\frac12,  \label{con5}
\end{equation}
where the integrated polarized quark densities $\Delta U=-0.2$ and
$\Delta S=0.6$ for the $\Lambda$ can be extracted by using SU(3)
symmetry from the deep-inelastic lepton-proton scattering
experiment data~\cite{SMC95} and the hyperon semileptonic decay
constants $F=0.459$ and $D=0.798$~\cite{Barnett96}. There might be
a large uncertainties in these values of $\Delta U$ and $\Delta
S$, since SU(3) symmetry breaking may affect the explicit
flavor-dependent helicity separation of the octet baryons
\cite{Ehr95}. Nevertheless, the effect of these uncertainties on
the pQCD fragmentation functions in the medium and large $z$
region, which give the main contributions to the spin transfers,
do not change the qualitative features of our results due to the
helicity retention property of the pQCD analysis. The five
constraints in (\ref{con1})-(\ref{con5}) leave us with three
unknown parameters, which are chosen to be $\tilde{A}_s$,
$\tilde{C}_s$ and $\tilde{C}_u$. Following Ref.~\cite{MSY4}, we
let them be the same with the value of 2. The $d$ quark
distributions are the same as those for the $u$ quark.

\subsection{Fragmentation functions via Gribov-Lipatov relation}

Unfortunately, we cannot directly measure the above described
quark distributions of the  $\Lambda$, since it is not possible to
use the $\Lambda$ as a target due to its short life time. Also one
obviously cannot produce a beam of charge-neutral $\Lambda$s. What
one can observe with  experiments is the quark to $\Lambda$
fragmentation, and therefore one needs a relation between  the
quark fragmentation and distribution functions. In order to
connect the fragmentation functions with the distribution
functions, we use~\cite{MSSY} the Gribov-Lipatov (GL)
relation~\cite{GLR}
\begin{equation}
D_q^h(z) \sim z \, q_h(z), \label{GLR}
\end{equation}
where $D_q^h(z)$ is the fragmentation function for a quark $q$
splitting into a hadron $h$ with longitudinal momentum fraction
$z$, and $q_h(z)$ is the quark distribution of finding the quark
$q$ inside the hadron $h$ carrying a momentum fraction $x=z$. The
GL relation is only known to be valid near $z \to 1$ in an energy
scale $Q^2_0$ in leading order approximation~\cite{BRV00}.
However, with this relation, predictions of $\Lambda$
polarizations~\cite{MSSY} based on quark distributions of the
$\Lambda$ in the SU(6) quark diquark  spectator model and in the
pQCD based counting rules analysis, have been found to be
supported by all available data from longitudinally polarized
$\Lambda$ fragmentation in
$e^+e^-$-annihilation~\cite{ALEPH96,DELPHI95,OPAL97}, polarized
charged lepton DIS process~\cite{HERMES,E665}, and most recently,
neutrino (antineutrino) DIS process~\cite{NOMAD}. Thus we still
use (\ref{GLR}) as an ansatz to relate the quark fragmentation
functions for the $\Lambda$ to the corresponding quark
distributions.

\section{Spin transfer for $\Lambda$ production in $\vec{p}p$ collisions}

The spin transfer for the produced $\Lambda$ in $\vec{p}p$
collisions is  mainly  determined by three subprocesses. The cross
section of the most important subprocess $q g \to q g$ strongly
depends on  the quark $q$  distribution in the colliding protons. The
strange quark contribution to the spin transfer to the $\Lambda$
is suppressed due to the fact that the strange quark is not
a valence quark  of the proton. As opposed to the $e^+e^-$
annihilation process where the $\Lambda$ polarization is dominated by
the strange quark  fragmentation,  $\vec{p}p$ collisions should
be a suitable place to check the $u$ and $d$ quark fragmentation
functions by measuring the large rapidity dependence of the spin
transfer to the $\Lambda$. In order to show the dominant  quark
contributions, the $u$ quark to  $\Lambda$
fragmentation functions  in the pQCD analysis and the SU(6) diquark model
are shown in the left-upper part of Fig.~\ref{Fig4}.

\begin{figure}
\begin{center}
\leavevmode {\epsfysize=14cm \epsffile{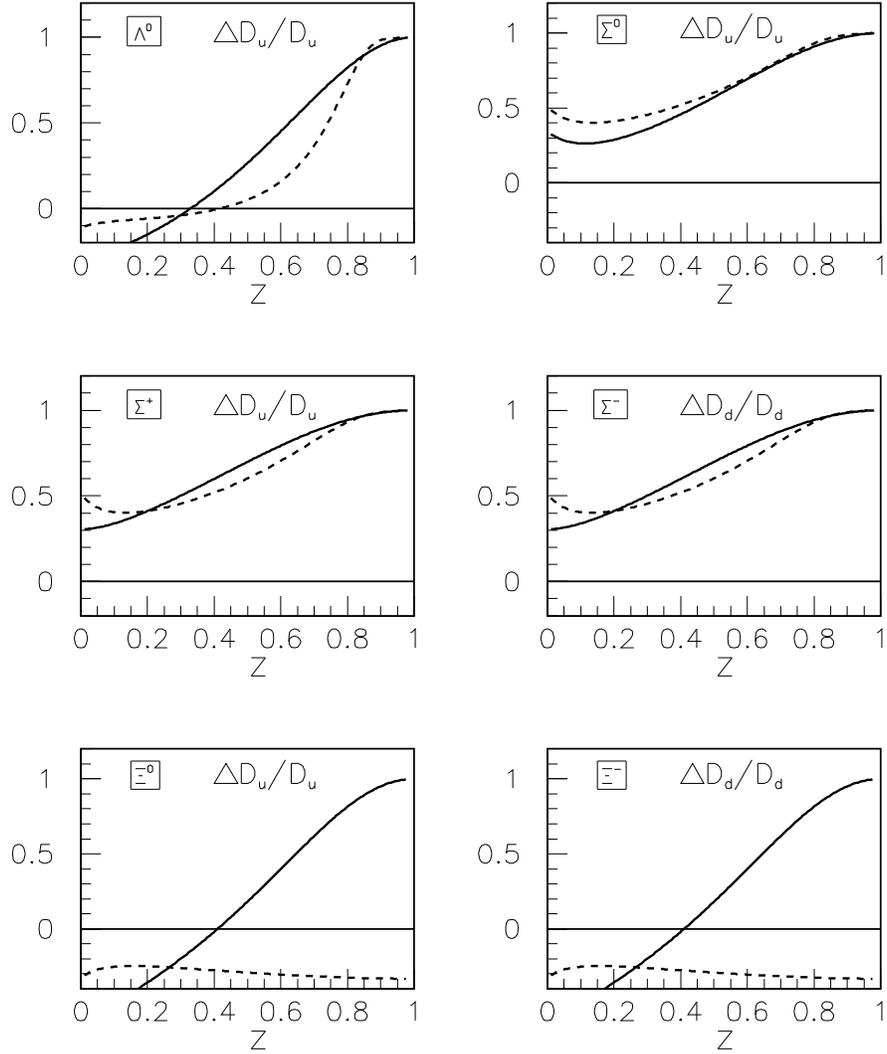}}
\end{center}
\caption[*]{\baselineskip 13pt The ratios of polarized to
unpolarized fragmentation functions for non-strange quarks, both
in the pQCD counting rules analysis (solid curves) and the SU(6)
quark-diquark spectator model (dashed curves). }\label{Fig4}
\end{figure}

By inserting the fragmentation functions obtained in the pQCD
analysis and the SU(6) quark diquark model into (\ref{A}), and
taking the minimal cutoff of the transverse momentum $p_T=13
~\rm{GeV}$, we obtain the spin transfers for $\Lambda$ production
in polarized $pp$ collisions at $\sqrt{s}=500~\rm{GeV}$. The
results are shown in the left-upper part of Fig.~\ref{Fig5A}.
We also show in Fig.~\ref{Fig5B} the results at
$\sqrt s = 200~\rm{GeV}$, an energy at which RHIC-BNL will
be operating. As expected, the spin transfer is a bit larger than
at $\sqrt s =500~\rm{GeV}$, since the cross section is smaller.

In our numerical calculations, we adopt the LO set of unpolarized
parton distributions of Ref.~\cite{GRV95} and polarized parton
distributions of LO GRSV standard scenario~\cite{GRSV96}. The spin
transfers as a function of the transverse momentum of the produced
$\Lambda$ in $\vec{p}p$ collisions at $\sqrt{s}=500~\rm{GeV}$,
with the specified values of its rapidity $y=0$ and $y=2$, are
given in Fig.~\ref{Fig6} (a) and (b) respectively. We find that
one should measure the spin transfers in the large $p_T$ region
for $y=0$ and select the events with $p_T \sim 30 ~\rm{GeV}$ for
$y=2$ in order to distinguish between different sets of
fragmentation functions in the two models.

\begin{figure}
\begin{center}
\leavevmode {\epsfysize=14cm \epsffile{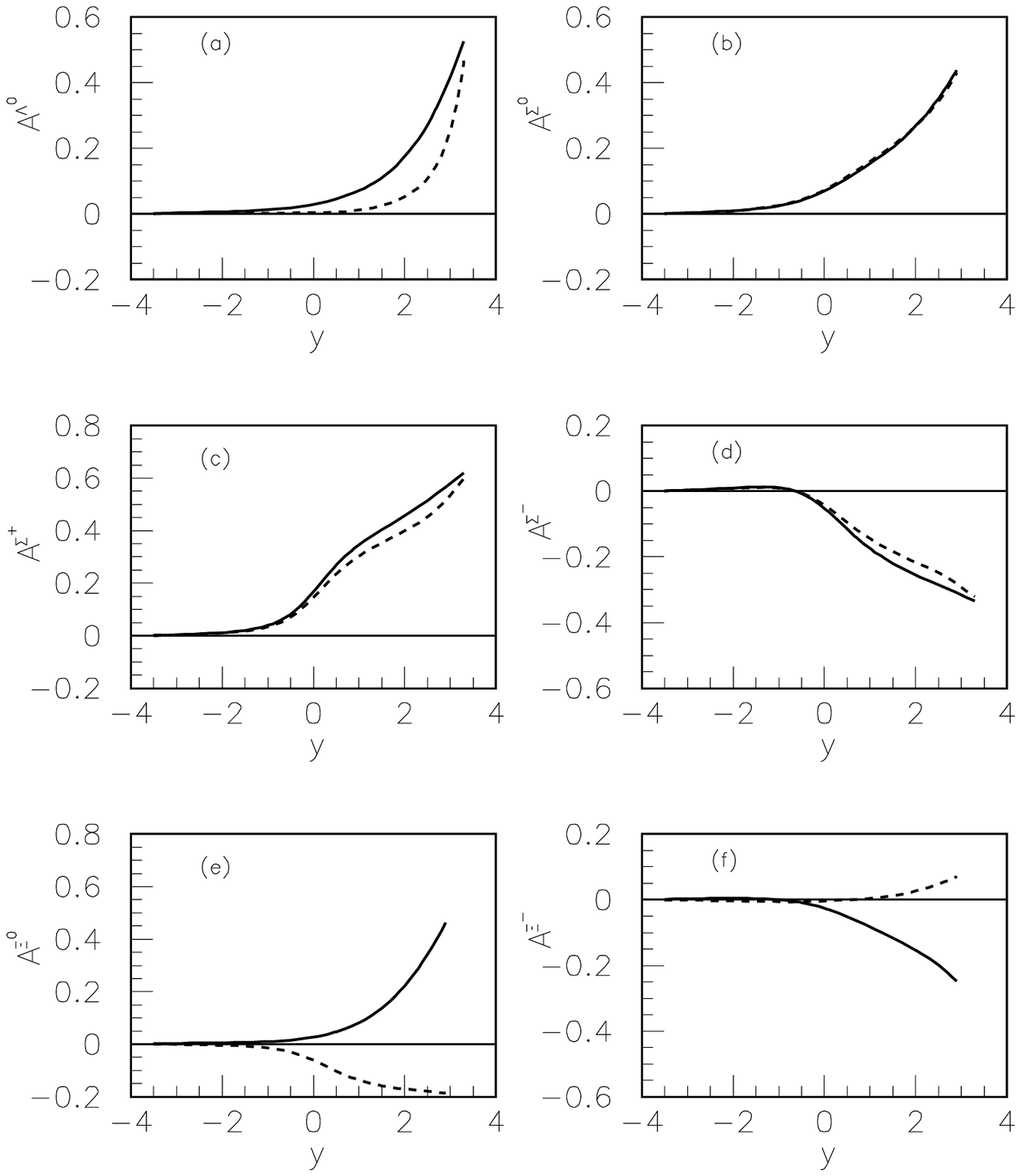}}
\end{center}
\caption[*]{\baselineskip 13pt The spin transfers as functions of
rapidity of the produced $\Lambda$, $\Sigma$, $\Xi$ of octet
baryon members, in $\vec{p}p$ collisions at
$\sqrt{s}=500~\rm{GeV}$, with the spin-dependent fragmentation
functions in  the pQCD counting rules analysis (solid curves) and
the SU(6) quark-diquark spectator model (dashed curves). Note that
the dashed and solid curves in (b) almost overlap.}\label{Fig5A}
\end{figure}

\begin{figure}
\begin{center}
\leavevmode {\epsfysize=14cm \epsffile{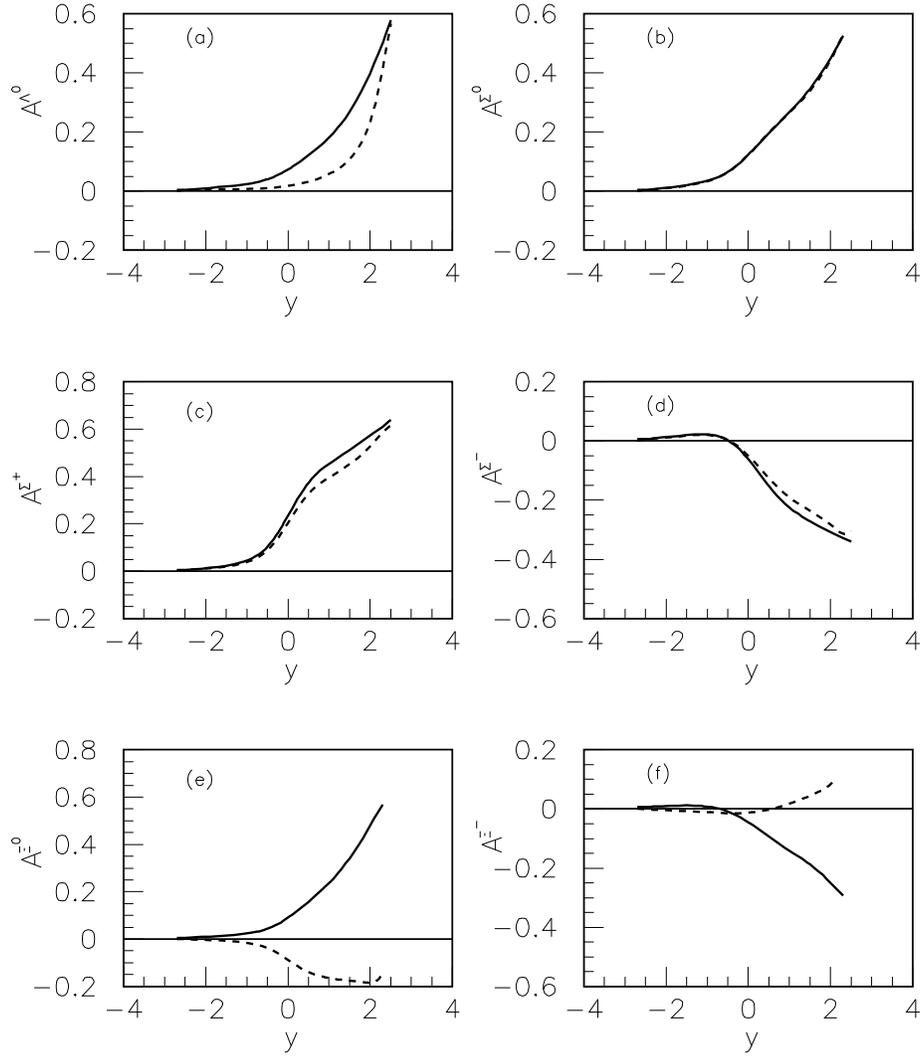}}
\end{center}
\caption[*]{\baselineskip 13pt The same as Fig.~\ref{Fig5A}, but
for $\vec{p}p$ collisions at $\sqrt{s}=200~\rm{GeV}$.}\label{Fig5B}
\end{figure}

\begin{figure}
\begin{center}
\leavevmode {\epsfysize=5cm \epsffile{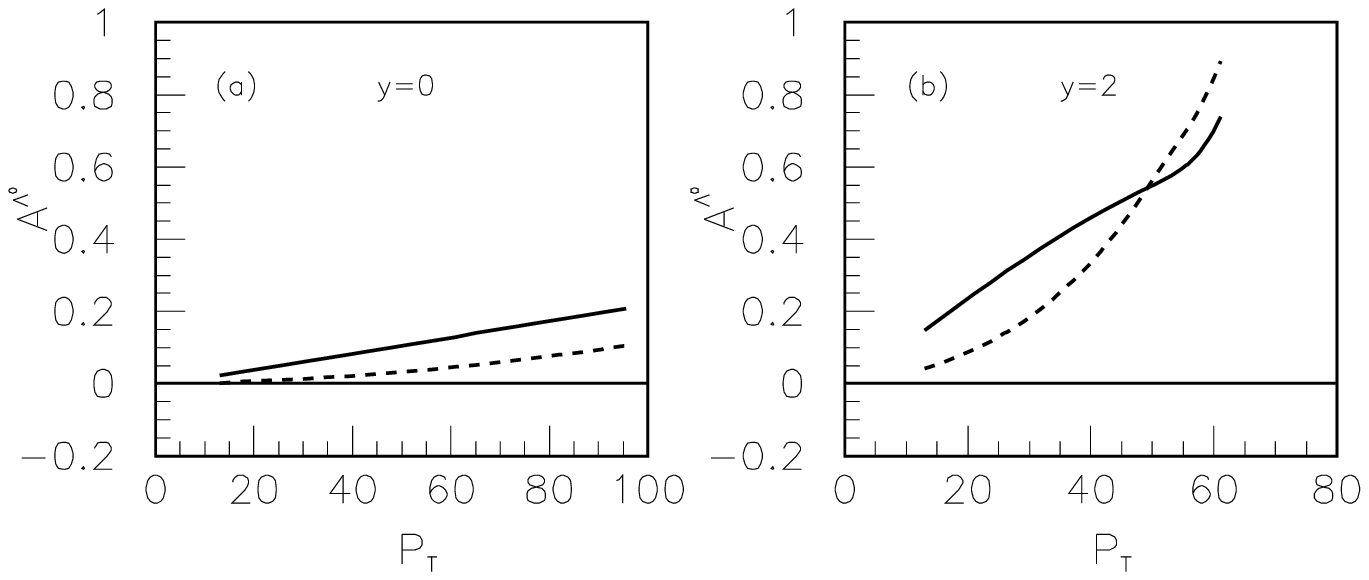}}
\end{center}
\caption[*]{\baselineskip 13pt The spin transfers as functions of
the transverse momentum of the produced $\Lambda$ in $\vec{p}p$
collisions at $\sqrt{s}=500~\rm{GeV}$, with for the fragmentation
functions in the pQCD counting rules analysis (solid curves) and
the SU(6) quark-diquark spectator model (dashed curves).
}\label{Fig6}
\end{figure}

Now let us make  a comparison  between our results and those in
Refs.~\cite{FSVb,BLTh}. In Ref.~\cite{FSVb}, Florian, Stratmann
and Vogelsang made  their predictions within  three different
scenarios of the polarized fragmentation functions. Scenario 1
corresponds to the SU(6) symmetric non-relativistic quark model,
according to which the whole $\Lambda$ spin is carried by the $s$
quark. Scenario 2 is based on an SU(3) flavor symmetry analysis
and on the first moment of $g_1$, and leads to the prediction that
the $u$ and $d$ quarks of the $\Lambda$ are negatively polarized.
Scenario 3 is built on the assumption that all light quarks
contribute equally to the $\Lambda$ polarization. It is very
interesting that the best agreement with available LEP data was
obtained within scenario 3~\cite{FSVa}, {\it i.e.} the $u$ and $d$
quark fragmentation functions are positively polarized. Our
predictions in the SU(6) quark diquark model and the pQCD analysis
are similar to those with scenario 3 polarized fragmentation
function in Ref.~\cite{FSVb}. In addition, our predictions are
also close to those predicted in Ref.~\cite{BLTh}, where positive
polarized  $u$ and $d$ quark fragmentation functions were used.
After all, our analysis shows  that the $u$ and $d$ quarks to
$\Lambda$ fragmentation functions are positively polarized at
least at large $z$, which is consistent with the results we found
in other processes~\cite{MSSY}.

\section{Extension of the analysis to other octet baryons}

In addition to the measurement of polarization of the produced
$\Lambda$, the detection technique of $\Sigma$ and $\Xi$ hyperons
is  also getting  more and more mature in order to measure the
various quark to hyperon fragmentation functions
\cite{SigmaP,XiP,Sigma0P}. Except for the $\Sigma^0$, which decays
electromagnetically, all other hyperons in the octet baryons have
their major decay modes mediated by weak interactions. Because
these weak decays do not conserve parity, information from their
decay products can be used to determine their polarization
\cite{SigmaP,XiP}. The polarization of $\Sigma^0$ can be also
re-constructed from the dominant decay chain $\Sigma^0 \to \Lambda
\gamma$ and $\Lambda \to p \pi^-$ \cite{Sigma0P}. Therefore we can
use the measurable fragmentation functions in order to extract
information on the spin and flavor content of hyperons, using the
available experimental facilities. Hence, it is important to
extend the analysis for the  $\Lambda$ to other octet baryons.
This can be done directly by adopting the same parameters for the
SU(6) quark-diquark model and pQCD analysis as those in
Ref.~\cite{MSY4}. In Fig.~\ref{Fig4}, we show the ratios of
non-dominant quark polarized to to unpolarized fragmentations into
octet baryons $\Sigma^0$, $\Sigma^+$, $\Sigma^-$, $\Xi^0$ and
$\Xi^-$. The spin transfers as functions  of the  rapidity of the
produced octet baryons in $\vec{p}p$ collisions,  at
$\sqrt{s}=500~\rm{GeV}$, and for the spin-dependent fragmentation
functions, both  in the pQCD counting rules analysis (solid lines)
and the SU(6) quark-diquark spectator model (dashed lines), are
presented in Fig.~\ref{Fig5A}.  For completeness the results
at $\sqrt s = 200~\rm{GeV}$ are shown in Fig.~\ref{Fig5B}.
By comparing the spin transfers in
Fig.~\ref{Fig5A} and Fig.~\ref{Fig5B} with the corresponding spin structure of the
fragmentation functions in Fig.~\ref{Fig4}, we find that the spin
transfers at large $y$ are mainly related to the non-dominant $u$ and $d$
quark fragmentation ratios of polarized to unpolarized
fragmentation functions at medium $z$ values. This can explain
qualitatively the different predictions of the spin transfers in
different models. The predictions for the spin transfers in the
two models are qualitatively similar for $\Lambda$ and $\Sigma$,
as can be seen from Fig.~\ref{Fig5A}(a)-(d). However, we find that
the spin transfers for the produced $\Xi$ can provide more clear
information to distinguish between the SU(6) quark diquark model
and pQCD based analysis. Hence, the $\Xi$ polarizations in
$\vec{p}p$ collisions deserve experimental attention.

In order to complete our  analysis, we include the spin transfers
as functions of the  rapidity of the produced proton and neutron
in $\vec{p}p$ collisions at $\sqrt{s}=500~\rm{GeV}$. The results
are given in Fig.~\ref{Fig7}. As shown in Fig.~\ref{Fig7}(b), the
spin transfer for the produced neutron is also a suitable quantity
for distinguishing the two sets of fragmentation functions  in the
two different models, but experimentally it is difficult to
measure the polarization of a fast neutron or that of a fast
proton.

\begin{figure}
\begin{center}
\leavevmode {\epsfysize=5cm \epsffile{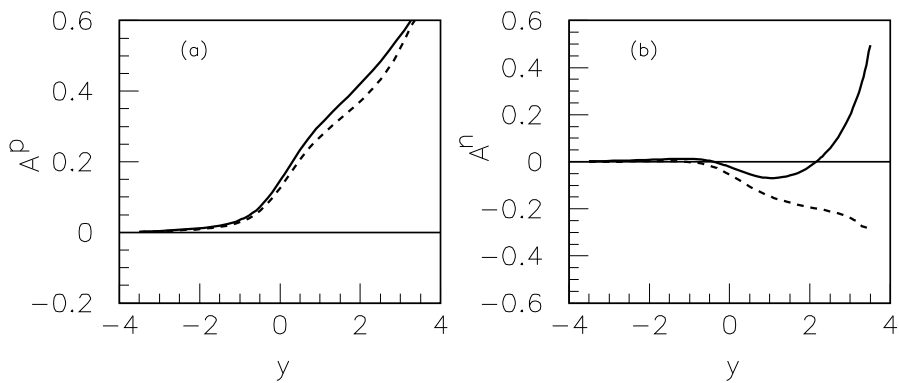}}
\end{center}
\caption[*]{\baselineskip 13pt
The same as Fig.~\ref{Fig5A}, but for the produced proton (a) and neutron (b).
}\label{Fig7}
\end{figure}

\section{An approximate estimate of spin transfers}

Usually it is hard to extract exact information on the inclusive
production of longitudinally polarized baryons in  $pp$ collisions
because of the following three complex aspects of the spin
transfer: (1) Many subprocesses are involved; (2) The contribution
of every subprocess includes four factors, {\it i.e.} the quark
helicity distributions of the proton, the polarization of the
produced baryon fragmentation functions, and the subprocess cross
sections $\Delta \hat{\sigma}$ and $\hat{\sigma}$; (3)  The
kinematic variables are integrated over.  In order to extract some
useful information from the above complex situation, we focus our
attention on the dominant subprocess, {\it i.e.} $q g \to q g$. We
can use the mean value theorem and take the cross sections out of
the corresponding integrals in both the numerator and denominator
in the expression for the spin transfer. Fortunately, $\Delta
\hat{\sigma}/ \hat{\sigma}$ for this subprocess is equal to one
for all $y$ (see Fig.~\ref{Fig3}). So the contribution of this
subprocess to the spin transfer only comes from two factors, {\it
i.e.} the quark helicity distributions and the polarization of the
fragmentation functions. In order to pin down  the roles played by
the above two factors in the spin transfer, we go back to review
the results shown in Fig.~\ref{Fig1} and Fig.~\ref{Fig2}. From
Fig.~\ref{Fig1}, one can see  that
there is a symmetry for $ y \leftrightarrow -y$ in the unpolarized
cross sections. However, there is  a strong asymmetry in the
polarized cross sections,  as shown in Fig.~\ref{Fig2}. This asymmetry may
arise from the corresponding asymmetry between $\Delta f_a^p(x_a)$
and $f_b^p(x_b)$ when  $a \leftrightarrow b$, and another possible
source for this asymmetry is the asymmetry of $z$ when $y
\leftrightarrow -y$. By comparing Fig.~\ref{Fig2}(a) and (b), one can see
that the asymmetry in the polarized cross section is mainly due to
the asymmetry between $\Delta f_a^p(x_a)$ and $f_b^p (x_b)$ when
$a \leftrightarrow b$ since the polarized cross section asymmetry
still remains  when $\Delta D_q^B$ is removed. In addition, we
find that the magnitude  of the polarized cross section in
Fig.~\ref{Fig2}(b) can be approximately obtained from that in
Fig.~\ref{Fig2}(a) by
multiplying it with a factor of $\Delta D_q^\Lambda$ at $z \simeq
0.65$. It turns out that the asymmetry in the polarized cross
section mainly comes from the helicity distributions of the proton
and the magnitude of the cross section is related to the
fragmentation functions at $z \simeq 0.65$.

\begin{figure}
\begin{center}
\leavevmode {\epsfysize=5cm \epsffile{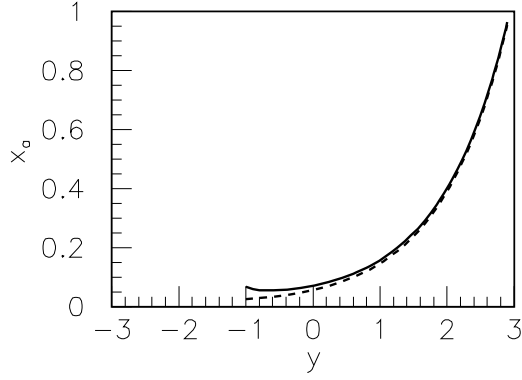}}
\end{center}
\caption[*]{\baselineskip 13pt $x_a$ as a function of $y$ for
$p_T=17 ~\rm{GeV}$, $z=0.65$ and $\sqrt{s}=500~\rm{GeV}$. The
solid and dashed curves correspond to $x_b=0.2$ and $x_b=0.7$,
respectively. }
\label{Fig8}
\end{figure}

\begin{figure}
\begin{center}
\leavevmode {\epsfysize=3.6cm \epsffile{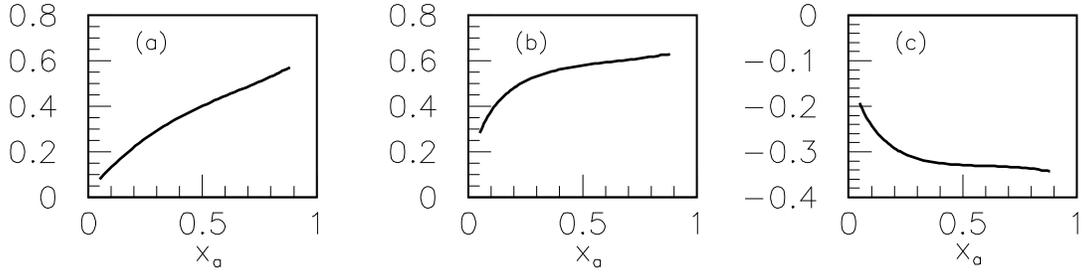}}
\end{center}
\caption[*]{\baselineskip 13pt
The quark helicity distributions of the proton.
(a) $(\Delta u +\Delta d)/(u +d)$;
(b) $\Delta u /u $;
(c) $\Delta d/d$; for the LO set of
unpolarized parton distributions of Ref.~\cite{GRV95}
and the polarized parton distributions of the LO
GRSV standard scenario~\cite{GRSV96} at $Q^2=10~\rm{GeV}^2$.}
\label{Fig9}
\end{figure}

Now let us show why there is a very strong asymmetry in the
$y$-dependence of the spin transfer. As an approximation, we only
consider the dominant subprocess $q g \to q g$, the spin transfer
can be expressed in the simple form

\begin{equation}
 A^B= \frac{\Delta q (x_a) g (x_b) \Delta D_q^B(z) + \Delta g (x_a) q (x_b)
 \Delta D_g^B (z)}
 { q (x_a) g (x_b) D_q^B(z) +  g (x_a) q (x_b) D_g^B (z)}.
 \end{equation}
The quark  distributions and fragmentation functions in this
expression should be understood in an average sense. The $g \to B$
fragmentation functions are much less known than the $q \to B$
fragmentation functions, and $\Delta D_g^B$ is customarily set to
zero at an initial energy scale. So for the moment let us neglect
the $g \to B$ fragmentation functions. Actually, this is what we
have done in the above exact calculation of $A^B$. Therefore, for
the $\Lambda$, we find an approximate formula

  \begin{equation}
  A^B= \left [ \frac{\Delta u + \Delta d}{u +d} \right ] (x_a)
  \left [ \frac{\Delta D_u^\Lambda}{D_u^\Lambda} \right ] (z \simeq 0.65). \label{AL}
  \end{equation}
In this expression there is a  $y$-dependence in  $x_a$  via

\begin{equation}
x_a=\frac{x_b p_T e^{y}}{x_b z\sqrt{s}-p_T e^{-y}}
\end{equation}
by setting  $p_T=17 ~\rm{GeV}$, $z=0.65$, $\sqrt{s}=500~\rm{GeV}$
and $x_b$  in the range  [0.2,  0.7]. In Fig.~\ref{Fig8}, we show $x_a$ as
a function of $y$ and the two curves indicate that the
$y$-dependence of $x_a$ is stable when  $x_b$ varies in the range
[0.2,  0.7]. The important feature we notice in Fig.~\ref{Fig8} is that
there is a strong asymmetry in $x_a$ when $ y \leftrightarrow -y$,
{\it i.e.} for a given $|y|$, the value of $x_a$ for $y<0$ is much
lower than that for $y>0$. In order to see how this asymmetry is
reflected in  the asymmetry in the spin transfer, we show the
helicity distributions of the proton in Fig.~\ref{Fig9}. For the $\Lambda$
case, we only need Fig.~\ref{Fig9}(a). Figs.~\ref{Fig9}(b)-(c)
will be used for the
other octet baryons. From Fig.~\ref{Fig9}(a), we find that the ratio
($\Delta u + \Delta d)/(u+d)$ increases with $x_a$. By combining
information from Fig.~\ref{Fig8}, Fig.~\ref{Fig9}(a), and Eq.~(\ref{AL}), the
asymmetry for the approximate formula, as shown in Fig.~\ref{Fig10}(a), can
be easily understood.

On the other hand, by looking at the results in Fig.~\ref{Fig5A}, we observe
that  Fig.~\ref{Fig5A}(a) and (b) are similar, (c) and (d) are mirror
symmetric, (e) and (f)  are  almost mirror symmetric, which
motivates us to extend the approximate estimate from the $\Lambda$
to the other octet baryons. For Fig.~\ref{Fig5A}(b) we use the same
formula as for (a). Similarly, we approximate Fig.~\ref{Fig5A}(c) and (e)
with $[{\Delta u}/u] (x_a) [ {\Delta D_u^B}/D_u^B](z\simeq ~0.65)$
and for Fig.~\ref{Fig5A}(d) and (f), we  replace the $u$ quark  by
the $d$ quark. It means that the asymmetry allows only to test the
ratio of polarized to unpolarized fragmentation functions of the
dominant quark in a region where $z \simeq 0.65$. The asymmetry is
mainly driven by the corresponding quark helicity asymmetries of
the $u$ and $d$ quark distributions in the proton. Our approximate
formulae  for all octet baryons are shown in Table 1, and the
approximate $y$-dependence of the spin transfers is shown in
Fig.~\ref{Fig10}. We find that the spin transfers are well described by our
approximate formulae in the given region of $y$.

\centerline{Table~1~~ The approximate formulae for the spin
transfers to octet baryons }

 \vspace{0.3cm}

\begin{Large}
\begin{center}
\begin{tabular}{|c||c|}\hline
Baryon& Approximate Formula for $A^B$\\ \hline
~~~~p~~~~&
$\frac{[\Delta u(x_a) \Delta D_u^p(z)+ \Delta d(x_a) \Delta D_d^p(z) ]}
{[u(x_a) D_u^p(z)+ d(x_a) D_d^p(z) ]}
$\\ \hline
~~~~n~~~~ &
$\frac{[\Delta u(x_a) \Delta D_u^n(z)+ \Delta d(x_a) \Delta D_d^n(z) ]}
{[u(x_a) D_u^n(z)+ d(x_a) D_d^n(z) ]}
$\\ \hline
$~~~~\Sigma^{+}~~~~$ &
$ \frac{\Delta u(x_a)}{u(x_a)}
\frac{\Delta D_u^{\Sigma^+}(z)}{D_u^{\Sigma^+} (z)}
$ \\ \hline
$~~~~\Sigma^{0}~~~~$ &
$\frac{(\Delta u(x_a)
+\Delta d(x_a))}{(u(x_a)+d(x_a))}
\frac{\Delta D_u^{\Sigma^0}(z)}{ D_u^{\Sigma^0} (z)}
$\\ \hline
$~~~~\Sigma^{-}~~~~$ &
$\frac{\Delta d(x_a)}{d(x_a)}
\frac{\Delta D_d^{\Sigma^-}(z)}{ D_d^{\Sigma^-} (z)}
$ \\ \hline
$~~~~\Lambda^{0}~~~~$ &
$ \frac{(\Delta u(x_a)
+\Delta d(x_a))}{(u(x_a)+d(x_a))}
\frac{\Delta D_u^\Lambda(z)}{ D_u^\Lambda (z)}
$ \\ \hline
$~~~~\Xi^{-}~~~~$ &
$\frac{\Delta d(x_a)}{d(x_a)}
\frac{\Delta D_d^{\Xi^-}(z)}{ D_d^{\Xi^-} (z)}
$ \\ \hline
$~~~~\Xi^{0}~~~~$ &
$\frac{\Delta u(x_a)}{u(x_a)}
\frac{\Delta D_u^{\Xi^0}(z)}{ D_u^{\Xi^0} (z)}
$ \\ \hline
\end{tabular}
\end{center}
\end{Large}

\begin{figure}
\begin{center}
\leavevmode {\epsfysize=14cm \epsffile{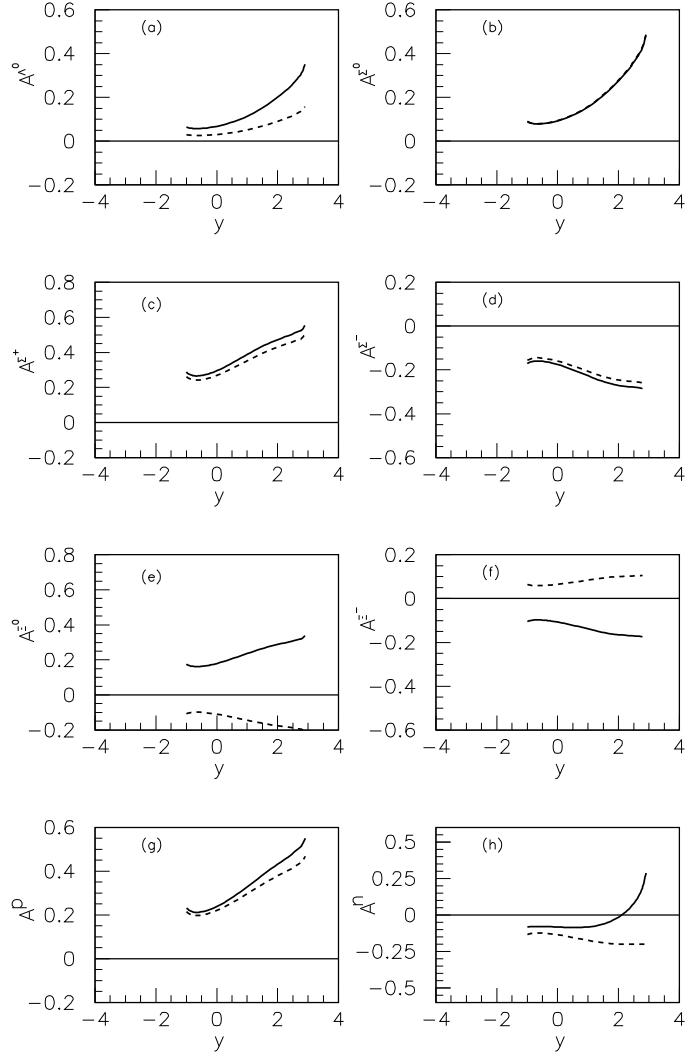}}
\end{center}
\caption[*]{\baselineskip 13pt The spin transfers as functions of
the rapidity of the produced  octet baryons  in $\vec{p}p$
collisions at $\sqrt{s}=500~\rm{GeV}$. The thin curves
are obtained using the approximate formulae in Table 1 and
the thick curves are the exact calculations.
The solid and dashed curves correspond to the
results with the fragmentation functions in  the pQCD counting
rules analysis  and the SU(6) quark-diquark spectator model,
respectively. Note that the dashed and solid curves in (b) almost
overlap. }\label{Fig10}
\end{figure}

The spin transfer in the positive $y$ region depends mainly on the
helicity distributions in the proton. This is because our present
knowledge about the gluon to a  octet baryon fragmentation
function is very poor and usually the polarized gluon
fragmentation functions are set to  zero at an initial energy
scale and they are only produced via  QCD evolution. There is a
strong suppression of the spin transfer in the negative $y$
region, due to the smallness of the quark helicity distribution
for small {\it x}. If the gluon fragmentation functions have a
significant polarization, then the spin transfer in the negative
$y$ region would not be zero although it would be  much smaller
than in the positive $y$ region.

By means of the approximate formulae for the $\Lambda$, we can
immediately check  whether the $u$ quark fragmentation is
positively  or negatively  polarized, according to the measured
results of the spin transfer. If the spin transfers are positive,
then the $u$ and $d$ fragmentation functions should be positively
polarized at large $z$ ($\Delta D_u^{\Lambda} (z) > 0$), and {\it
vice versa}. With our approximate formula the results of
Refs.~\cite{FSVb} and \cite{BLTh} can also be easily understood.

\section{Summary}

In summary, we have considered the inclusive production of
longitudinally polarized baryons in ${\vec p}p$ collisions at
RHIC-BNL, with one longitudinally polarized proton. We predicted
the spin transfer between the initial proton and the produced
baryon as a function of its rapidity by means of an SU(6) quark
diquark spectator model and a pQCD analysis. The same analysis was
extended from the $\Lambda$ case to other octet baryons. We found
that three subprocesses including $q g \to q g$, $ q q \to q q $
and $q q^\prime \to q q^\prime$ have dominant contributions to the
spin transfers. We pointed out some sensitive kinematics regions
where one can distinguish between different sets of fragmentation
functions. We tried some approximate formulae to describe the spin
transfer and found that the asymmetry allows only to test the
ratio of polarized to unpolarized fragmentation functions of the
dominant quark in a region where $z=0.65$ and it is mainly driven
by the corresponding quark helicity asymmetries of the $u$ and $d$
quarks in the proton. Our predictions for the spin
transfers with positively polarized $u$ and $d$ quark
fragmentations to $\Lambda$ should be checked soon by the RHIC-BNL
experimental data, and if the measurements for other produced
octet baryons in $\vec{p}p$ collisions can be realized, they will
enrich our knowledge of the hadronization mechanism.

{\bf Acknowledgments: } We are grateful to Werner Vogelsang for
providing the code they have used for $\Lambda$ production in
Ref.~\cite{FSVb}. This work is
partially supported by National Natural Science Foundation of
China under Grant Numbers 19975052, 10025523 and 19875024, by Fondecyt
(Chile) 3990048, by the cooperation
programmes Ecos-Conicyt C99E08 between France and
Chile, and by Fondecyt (Chile) grant 8000017, and by
a C\'atedra Presidencial (Chile).

\end{document}